\documentclass[11pt,a4paper]{article}
\usepackage[mathcal]{eucal}
\usepackage{latexsym}
\usepackage{amssymb}
\begin{titlepage}
\title{Nonlinear gravitons from the initial value constraints of GR in Ashtekar variables}
\author{Eyo Eyo Ita III}
\medskip
\input amssym.def
\input amssym.tex

\def \in{\indent}

\begin{document}
\maketitle
\bigskip
\centerline{Department of Applied Mathematics and Theoretical Physics} 
\smallskip
\centerline{Centre for Mathematical Sciences, University of Cambridge, Wilberforce Road}
\smallskip
\centerline{Cambridge CB3 0WA, United Kingdom}
\smallskip
\centerline{eei20@cam.ac.uk} 
  
\bigskip
    
\begin{abstract}
In this paper we provide a possible realization of Penrose's idea of nonlinear gravitons by constructing a solution to the initial value constraints in Ashtekar variables.  The solution inputs are a spatial SU(2) connection and two free functions of position, and can be constructed as a formal operatorial expansion in powers of the cosmological constant about spacetimes of Petrov Type O.  We first present the linear case, and then provide a simple nonlinear example to first order using a spatially homogeneous connection.
\end{abstract}
\end{titlepage}
  
\section{Introduction: The linear graviton}
In \cite{PENROSE} Roger Penrose takes issue with the standard view of the graviton as a weak-field perturbation of a background spacetime.  He proposes the idea that each graviton should carry its measure of curvature, corresponding to a solution of the full nonlinear Einstein equations.  Motivated by Penrose's idea, in this paper we will provide a possible realization of the concept of nonlinear gravitons using the initial value constraints of general relativity (GR).  The initial value constraints of GR with cosmological constant $\Lambda$ in the Ashtekar variables \cite{ASH1} are the Gauss' law, diffeomorphism and Hamiltonian constraints $(G_a,H_i,H)$ given respectively by  
\begin{eqnarray}
\label{DEEEF}
D_i\widetilde{\sigma}^i_a=\partial_i\widetilde{\sigma}^i_a+f_{abc}A^b_i\widetilde{\sigma}^i_a=0;~~\epsilon_{ijk}\widetilde{\sigma}^j_aB^k_a=0;
~~\epsilon_{ijk}\epsilon^{abc}\widetilde{\sigma}^i_a\widetilde{\sigma}^j_b\Bigl({\Lambda \over 3}\widetilde{\sigma}^k_c+B^k_c\Bigr)=0,
\end{eqnarray}
\noindent
where $A^a_i$ is a left-handed $SU(2)$ gauge connection and $\widetilde{\sigma}^i_a$ is a densitized triad.\footnote{Symbols from the beginning part of the Latin alphabet $a,b,c,\dots$ denote internal indices, and from the 
middle $i,j,k,\dots$ denote spatial indices.}  We have also defined $B^i_a={1 \over 2}\epsilon^{ijk}(\partial_jA^a_k-\partial_kA^a_j+f^{abc}A^b_jA^c_k)$ as the 
magnetic field of $A^a_i$, and $D_iv_a=\partial_iv_a+f_{abc}A^b_iv_c$ is the $SU(2)_{-}$ covariant derivative with structure constants $f_{abc}=\epsilon_{abc}$.  Define the Ansatz 
\begin{eqnarray}
\label{CEEANSATZ}
\widetilde{\sigma}^i_a=-\Bigl({3 \over \Lambda}\delta_{ae}+\epsilon_{ae}\Bigr)B^i_e,~~(\hbox{det}B)\neq{0},
\end{eqnarray}
\noindent
where $\epsilon_{ae}\in{SU}(2)\otimes{SU}(2)$.  Substitution of (\ref{CEEANSATZ}) into (\ref{DEEEF}) yields
\begin{eqnarray}
\label{WRITEAS}
B^i_eD_i\{\epsilon_{ae}\}=0;~~\epsilon_{dae}\epsilon_{ae}=0;~~\hbox{tr}\epsilon+{\Lambda \over 3}Var\epsilon+{{\Lambda^2} \over 3}\hbox{det}\epsilon=0,
\end{eqnarray}
\noindent
where we have defined $Var{M}=(\hbox{tr}M)^2-\hbox{tr}M^2$.  To obtain the third equation of (\ref{WRITEAS}) we have cancelled off a factor of $\hbox{det}B\neq{0}$ as well as any numerical pre-factors.  Defining vector 
fields $\textbf{v}_e=B^i_e\partial_i$ and the magnetic `helicity density matrix' $C_{be}=A^b_iB^i_e$, then the Gauss' law constraint $G_a$ can be written as
\begin{eqnarray}
\label{GAU}
B^i_eD_i\epsilon_{ae}=\textbf{v}_e\{\epsilon_{ae}\}+\bigl(f_{abf}\delta_{ge}+f_{ebg}\delta_{af}\bigr)C_{be}\Psi_{fg}\equiv\textbf{w}_e\{\epsilon_{ae}\}=0.
\end{eqnarray}
\noindent
Note for $\epsilon_{ae}=0$ that (\ref{WRITEAS}) is identically satisfied, which corresponds to a particular class of spacetimes of Petrov Type O since all three eigenvalues of $\epsilon_{ae}$ are equal.  Let us first consider the case of a linear gravitational wave propagating on these background spacetimes.  Choose $\vert\epsilon_{ae}\vert<<{1 \over \Lambda}$ as a perturbation and choose a connection 
$A^a_i=\delta^a_i\alpha+a^a_i$, where $\vert{a}^a_i\vert<<\alpha$.  Then we substitute this Ansatz into (\ref{WRITEAS}) and expand to linear order in $\epsilon_{ae}$ and $a^a_i$.  The magnetic field to linearised order is 
given by $B^i_e=\epsilon^{ijk}\partial_ja^e_k-\alpha{a}^i_e+\delta^i_e\bigl(\alpha^2+\alpha\hbox{tr}a\bigr)$.  Since (\ref{WRITEAS}) is already of linearised order in $\epsilon_{ae}$, then we need only expand $B^i_e$ and $C_{ae}$ to zeroth order in $a^a_i$, since all terms of the form $\epsilon\epsilon$, $\epsilon{a}$ and $aa$ are of second order.  So we only need $\textbf{v}_e=B^i_e\partial_i\sim\alpha^2\partial_e$ and $C_{ae}\sim\delta_{ae}\alpha^3$.  Note that using $C_{be}\propto\delta_{be}$ causes the terms in brackets in (\ref{GAU}) to drop out due to antisymmetry of the structure constants $f_{abc}$.  Then the linearized counterparts to (\ref{WRITEAS}) to zeroth order in $\Lambda$ reduce to
\begin{eqnarray}
\label{LINEAR}
\partial_e\epsilon_{ae}=0;~~\epsilon_{dae}\epsilon_{ae}=0;~~\hbox{tr}\epsilon=0.
\end{eqnarray}
\noindent
Equation (\ref{LINEAR}) states that the perturbation $\epsilon_{ae}$ is transverse, traceless and symmetric, namely that it corresponds to a linearized gravitational wave.
\section{Generalization to the full nonlinear case}
Having shown that the initial value constraints can produce linearized gravitons, we will now progress to the full nonlinear case.  The full Hamiltonian constraint can be written in the form
\begin{eqnarray}
\label{GJK}
\hbox{tr}\epsilon+\Lambda{I}^{aebf}\epsilon_{ae}\epsilon_{bf}+\Lambda^2E^{aebfcg}\epsilon_{ae}\epsilon_{bf}\epsilon_{cf}=0,
\end{eqnarray}
\noindent
where we have defined $I^{aebf}={1 \over 3}\bigl(\delta^a_b\delta^e_f-\delta^a_f\delta^b_e\bigr)$ and $E^{aebfcg}={1 \over 3}\epsilon^{abc}\epsilon^{efg}$.  
Defining $e^f_{ae}$ and $E^f_{ae}$ as an orthogonal basis of diagonal and off-diagonal symmetric matrix elements, let us write $\epsilon_{ae}$ in the matrix form
\begin{displaymath}
\epsilon_{ae}=e^f_{ae}\varphi_f+E^f_{ae}\Psi_f=
\left(\begin{array}{ccc}
\varphi_1 & \Psi_3 & \Psi_2\\
\Psi_3 & \varphi_2 & \Psi_1\\
\Psi_2 & \Psi_1 & \varphi_3\\
\end{array}\right)
.
\end{displaymath}
\noindent
Then defining $C_{[ae]}\equiv{C}_{ae}-C_{ea}$, the Gauss' law constraint (\ref{GAU}) is given by $e^f_{ae}\textbf{w}_e\{\varphi_f\}+E^f_{ae}\textbf{w}_e\{\Psi_f\}=0$ with matrix form
\begin{displaymath}
\left(\begin{array}{ccc}
\textbf{v}_1-C_{[23]} & -C_{32} & C_{23}\\
C_{31} & \textbf{v}_2-C_{[31]} & -C_{13}\\
-C_{21} & C_{12} & \textbf{v}_3-C_{[12]}\\
\end{array}\right)
\left(\begin{array}{c}
\varphi_1\\
\varphi_2\\
\varphi_3\\
\end{array}\right)
\end{displaymath}
\begin{displaymath}
+
\left(\begin{array}{ccc}
C_{22}-C_{33} & \textbf{v}_3-C_{12}+2C_{21} & \textbf{v}_2-2C_{31}+C_{13}\\
\textbf{v}_3-2C_{12}+C_{21} & C_{33}-C_{11} & \textbf{v}_1-C_{23}+2C_{32}\\
\textbf{v}_2-C_{31}+2C_{13} & \textbf{v}_1-2C_{23}+C_{32} & C_{11}-C_{22}\\
\end{array}\right)
\left(\begin{array}{c}
\Psi_1\\
\Psi_2\\
\Psi_3\\
\end{array}\right)
=0.
\end{displaymath}
\noindent
Assuming invertibility of the matrix of differential operators $E^f_{ae}\textbf{w}_e$, then we can define $\hat{J}_f^g=-(E^f_{af^{\prime}}\textbf{w}_{f^{\prime}})^{-1}(e^g_{ag^{\prime}}\textbf{w}_{g^{\prime}})$ as a propagator from the diagonal to the off-diagonal elements of $\epsilon_{ae}$.  This enables one formally to rewrite the Gauss' law constraint as an embedding map
\begin{eqnarray}
\label{MAKINGTHE1}
\Psi_f=\hat{J}_f^g\varphi_g\longrightarrow\epsilon_{ae}=\bigl(e^g_{ae}+E^f_{ae}\hat{J}_f^g\bigr)\varphi_f\equiv\hat{T}_{ae}^f\varphi_f=\hat{T}_{ae}^f[A]\varphi_f.
\end{eqnarray}
\noindent
Then defining $Q^{fg}\equiv{I}^{aebf}\hat{T}^f_{ae}\hat{T}^g_{bf}$ and $Q^{fgh}\equiv{E}^{aebfcg}\hat{T}^{f^{\prime}}_{ae}\hat{T}^{g^{\prime}}_{bf}\hat{T}^{h^{\prime}}_{cg}$ the Hamiltonian constraint on solutions to the Gauss' law constraint becomes
\begin{eqnarray}
\label{MAKINGTHE2}
\varphi_3=-\varphi_1-\varphi_2-\Lambda{Q}^{f^{\prime}g^{\prime}}\varphi_{f^{\prime}}\varphi_{g^{\prime}}
-\Lambda^2Q^{f^{\prime}g^{\prime}h^{\prime}}\varphi_{f^{\prime}}\varphi_{g^{\prime}}\varphi_{h^{\prime}}\equiv\Lambda{Q}[\vec{\varphi};A].
\end{eqnarray}
\noindent
This can be solved by fixed point iteration by defining a sequence $(\varphi_3)_{(n)}$ where $(\varphi_3)_{(0)}=-\varphi_1-\varphi_2$ and the recursion relation $(\varphi_3)_{(n+1)}=Q[\varphi_1,\varphi_2,(\varphi_3)_{(n)};A]$.  Then the full solution, if convergent, is given by $\varphi_3=\hbox{lim}_{n\rightarrow\infty}(\varphi_3)_{(n)}$.  So the prescription for writing down a solution starts by making a choice of connection $A^a_i$.  This defines the Gauss' law propagator $\hat{J}_f^g=\hat{J}_f^g[A]$ and the integrodifferential matrix operator for the embedding map $\hat{T}^a_{ef}=\hat{T}^a_{ef}[A]$.  Then one chooses two free functions $\varphi_1$ and $\varphi_2$, and constructs the solution 
\begin{eqnarray}
\label{DOW}
\epsilon_{ae}=\hat{T}^1_{ae}\varphi_1+\hat{T}^2_{ae}\varphi_2+\hat{T}^3_{ae}\varphi_3[\varphi_1,\varphi_2;A]=\epsilon_{ae}[\varphi_1,\varphi_2;A].
\end{eqnarray}
\noindent
In (\ref{DOW}) $\varphi_1$ and $\varphi_2$ constitute the two physical degrees of freedom, and the connection $A^a_i$ forms an input into $\hat{T}^f_{ae}$ and into $\varphi_3$ through the latter.
\section{Example: spatially homogeneous connection}
We will demonstrate using a spatially homogeneous connection $A^a_i$, with $\varphi_f=n_fe^{\vec{k}\cdot\vec{r}}$ and $\Psi_f=m_fe^{\vec{k}\cdot\vec{r}}$.  For simplicity, we have chosen $m_f$, $n_f$ and $k_f$ as numerical constants.  The following objects follow from $A^a_i$ 
\begin{eqnarray}
\label{MAKINGTHE3}
B^i_a=(\hbox{det}A)(A^{-1})^i_a;~~C_{ae}=(\hbox{det}A)\delta_{ae};~~\textbf{v}_a=(\hbox{det}A)(A^{-1})^i_a\partial_i,
\end{eqnarray}
\noindent
and the vector fields $\textbf{v}_a$ have the following action
\begin{eqnarray}
\label{MAKINGTHE31}
\textbf{v}_a\{\varphi_f\}=(\hbox{det}A)e_a\varphi_f;~~\textbf{v}_a\{\Psi_f\}=(\hbox{det}A)e_a\Psi_f,
\end{eqnarray}
\noindent
where we have defined $e_a\equiv(A^{-1})^i_ak_i$.
\noindent
After dividing by $(\hbox{det}A)\neq{0}$, then the Gauss' law constraint propagator $\hat{J}_f^g$ consists of c-numbers, given by
\begin{displaymath}
\hat{J}^g_f={1 \over 2}
\left(\begin{array}{ccc}
(e_1)^2(e_2e_3)^{-1} & -e_2(e_1)^{-1} & -e_3(e_2)^{-1}\\
-e_1(e_3)^{-1} & (e_2)^2(e_3e_1)^{-1} & -e_3(e_1)^{-1}\\
-e_1(e_2)^{-1} & -e_2(e_1)^{-1} & (e_3)^2(e_1e_2)^{-1}\\
\end{array}\right)
.
\end{displaymath}
\noindent
In this case the action of the Gauss' law propagators $\hat{J}_f^g$ is algebraic, due to the functional form of $\epsilon_{ae}$.  So (\ref{GAU}) is the same as
\begin{eqnarray}
\label{MAKINGTHE4}
m_1={1 \over 2}\Bigl[\Bigl({{(e_1)^2} \over {e_2e_3}}\Bigr)n_1-\Bigl({{e_2} \over {e_3}}\Bigr)n_2-\Bigl({{e_3} \over {e_2}}\Bigr)n_3\Bigr];\nonumber\\
m_2={1 \over 2}\Bigl[-\Bigl({{e_1} \over {e_3}}\Bigr)n_1+\Bigl({{(e_2)^2} \over {e_3e_1}}\Bigr)n_2-\Bigl({{e_3} \over {e_1}}\Bigr)n_3\Bigr];\nonumber\\
m_3={1 \over 2}\Bigl[-\Bigl({{e_1} \over {e_2}}\Bigr)n_1-\Bigl({{e_2} \over {e_1}}\Bigr)n_2+\Bigl({{(e_3)^2} \over {e_1e_2}}\Bigr)n_3\Bigr].
\end{eqnarray}
\noindent
Using (\ref{MAKINGTHE4}), after some algebra we have the following relations 
\begin{eqnarray}
\label{MAKINGTHE5}
Var\epsilon=2\bigl(n_1n_2+n_2n_3+n_3n_1-(m_1)^2-(m_2)^2-(m_3)^2\bigr)\nonumber\\
=\vert{e}\vert^2\biggl[{{n_1n_2} \over {(e_3)^2}}+{{n_2n_3} \over {(e_1)^2}}+{{n_3n_1} \over {(e_2)^2}}
-{1 \over 2}\Bigl({{e_1} \over {e_2e_3}}\Bigr)^2(n_1)^2-{1 \over 2}\Bigl({{e_2} \over {e_3e_1}}\Bigr)^2(n_2)^2
-{1 \over 2}\Bigl({{e_3} \over {e_1e_2}}\Bigr)^2(n_3)^2\biggr];\nonumber\\
\hbox{det}\epsilon=n_1n_2n_3+2m_1m_2m_3-n_1(m_1)^2-n_2(m_2)^2-n_3(m_3)^3=0
\end{eqnarray}
\noindent
where we have defined $\vert{e}\vert^2=(e_1)^2+(e_2)^2+(e_3)^2$.  Note that the determinant of $\epsilon_{ae}$ vanishes.  Substitution of (\ref{MAKINGTHE5}) into (\ref{GJK}) yields
\begin{eqnarray}
\label{FINER}
n_3=-n_1-n_2-\Lambda{Q}[\vec{n};\vec{e}];~~Q={1 \over 3}\sum_{f=1}^3I^{fgh}\Bigl[\Bigl({{\vert{e}\vert} \over {e_f}}\Bigr)^2n_gn_h-{1 \over 2}\Bigl({{\vert{e}\vert{e}_fn_f} \over {e_ge_h}}\Bigr)^2\Bigr],
\end{eqnarray}
\noindent
where $I^{fgh}=1$ for even and for odd permutations of $f\neq{g}\neq{h}$, and zero otherwise.  Equation (\ref{FINER}) is a quadratic algebraic equation, which can be solved in closed form for $n_3$ as a function of $n_1$ and $n_2$.  But the whole point is to use the fixed point iteration procedure, which applies in the general case where $A^a_i$ is not constant.  To first order in $\Lambda$ this is given by
\begin{eqnarray}
\label{MAKINGTHE6}
(n_3)_{(1)}=-(n_1+n_2)+{{\Lambda\vert{e}\vert} \over {3e_1e_2e_3}}\Bigl[-{1 \over 2}\bigl((e_3)^2+(e_1)^2\bigr)n_1n_1\nonumber\\
-{1 \over 2}\bigl((e_2)^2+(e_3)^2\bigr)n_2n_2+\bigl((e_1e_2)^2-(e_3)^2\vert{e}\vert^2\bigr)n_1n_2\Bigr].
\end{eqnarray}
\noindent
This can be repeated to any desired order, producing an expansion in powers of $\Lambda$ whose coefficients depend on $(n_1,n_2)$.  The full solution is 
\begin{eqnarray}
\label{MAKINGTHE7} 
\widetilde{\sigma}^i_a=-\Bigl({\Lambda \over 3}\delta_{ae}+\bigl(\hat{T}^1_{ae}n_1+\hat{T}^2_{ae}n_2+\hat{T}^3_{ae}n_3[n_1,n_2;\vec{e}]\bigr)e^{\vec{k}\cdot{\vec{r}}}\Bigr)(A^{-1})^i_e(\hbox{det}A).
\end{eqnarray}
\section{Discussion}
We have presented a formula for nonlinear gravitons by expansion about Type O spacetimes where $\hbox{det}B\neq{0}$.  The solution is fixed by two free 
functions $\varphi_1$ and $\varphi_2$ and a connection $A^a_i$, and can be developed as an formal operatorial expansion by fixed point iteration.  We propose this as a possible realization of Penrose's idea \cite{PENROSE} at a formal level. 

\end{document}